# Enantioselective optical gradient forces using 3D structured vortex light

**KAYN A. FORBES\* AND DALE GREEN**

*School of Chemistry, University of East Anglia, Norwich NR4 7TJ, United Kingdom*
*Corresponding author: k.forbes@uea.ac.uk*

**Here we highlight enantioselective optical gradient forces present in 3D structured optical vortex tweezing systems. One chiral force originates from the circular polarization of the light, while remarkably the other is independent of the input polarization - even occurring for unpolarized light – and is not present in 2D structured light nor propagating plane waves. This latter chiral sorting mechanism allows for the enantioselective trapping of chiral particles into distinct rings in the transverse plane through conservative radial forces.**

http://dx.doi.org/10.1364/OL.99.099999

Light conveys energy and momentum - the basis for optical manipulation of matter [1]. Optical trapping encapsulates techniques involving one or more lasers that enable the three-dimensional control of particles via non-contact optical forces [2–4]. The *non-conservative* radiation pressure force stemming from the transfer of optical linear momentum acts to push particles in the direction of propagation (often referred to as the scattering force, though in terms of photons it is absorption and non-forward scattering); the *conservative* optical gradient force causes particles to be attracted to regions of high intensity (which is in fact the forward elastic scattering of photons); optical binding forces exist *between* particles [5]. An important class of laser modes are structured modes [6,7], and more specifically those that carry orbital angular momentum (OAM), known as optical vortices or twisted light [8]. Optical tweezers using twisted light has seen much activity [9–11], predominantly because they can transfer their OAM and induce mechanical motion of trapped particles. This transfer of OAM to cause rotational motion is completely distinct from the gradient trapping force itself; optical angular momentum is conserved in the gradient force. As far as we are aware no optical gradient force trapping scheme has utilized the OAM in a direct manner, i.e., a non-mechanical way. The OAM of optical vortices is quantified in discrete units of $\ell\hbar$ where $\ell \in \mathbb{Z}$, and the sign of $\ell$ designates what direction the helical wavefront twists: for $\ell > 0$ it is left-handed, while $\ell < 0$ is right-handed. This chirality of optical vortices has witnessed a profusion of research activity, with the field recently surveyed [12]. The question addressed in this paper is whether the OAM, both its sign and magnitude, can influence the optical gradient trapping force when applied to Rayleigh-sized chiral particles. Previous studies looking at chiral optical trapping forces were specifically concerned with circularly polarized unstructured or plane wave light interacting with small chiral particles [13–16]. Being able to optically sort chiral molecular enantiomers using all-optical methods at the nanoscale would have a profound impact on the drug and pharmaceutical industry. Beyond natural molecules, chiral trapping forces have been employed in a number of different setups, predominantly involving chiral nanoparticles and plasmonic tweezing systems [17–27], though there have been some studies involving structured laser light [23,28–32].

**Conservative chirality-sorting forces**: Quantum electrodynamics (QED) [33] is employed to describe the light-matter interactions that correspond to optical gradient forces in this paper. Truncated to the dipole approximation, the Power-Zienau-Woolley (PZW) multipolar interaction Hamiltonian describes this coupling as [34]

$$H_{\text{int}}(\xi) = -\varepsilon_0^{-1}\mu_i(\xi)d_i^{\perp}(\boldsymbol{R}_\xi) - m_i(\xi)b_i(\boldsymbol{R}_\xi), \quad (1)$$

where $\mu_i(\xi)$ and $m_i(\xi)$ are the electric and magnetic dipole transition operators, respectively; $d_i^{\perp}(\boldsymbol{R}_\xi)$ and $b_i(\boldsymbol{R}_\xi)$ are the transverse (with respect to the Poynting vector) electric displacement field and magnetic field mode operators, respectively, acting on a particle $\xi$ at the location $\boldsymbol{R}_\xi$; Einstein summation of repeated tensor indices is assumed throughout, i.e. $a_i b_i = \boldsymbol{a} \cdot \boldsymbol{b}$. The first term in (1) is the electric dipole coupling (E1) and the second magnetic dipole coupling (M1). For input circularly-polarized Laguerre-Gaussian (LG) modes the electromagnetic free field expansion operators truncated to first-order in the paraxial parameter $kw_0$ may be given by [35,36]

$$d_i^{\perp}(\boldsymbol{r}) = \sum_{k,\sigma,\ell,p} \Omega \left[ \left\{ e_i^{(\sigma)} + \frac{i}{\sqrt{2}k}\left(\frac{\partial}{\partial r} - \frac{\ell\sigma}{r}\right)e^{i\sigma\phi}\hat{z}_i \right\} \right.$$
$$\left. \times f_{|\ell|,p}(r) a_{|\ell|,p}^{(\sigma)}(k\hat{z}) e^{i(kz+\ell\phi)} - H.c. \right], \quad (2)$$

and

$$b_i(\mathbf{r}) = \sum_{k,\sigma,\ell,p} \frac{\Omega}{\varepsilon_0 c} \left[ \left\{ \overline{b}_i^{(\sigma)} + \frac{1}{\sqrt{2}k}\left(\sigma \frac{\partial}{\partial r} - \frac{\ell}{r}\right) e^{i\sigma\phi} \hat{z}_i \right\} \right.$$
$$\left. \times f_{|\ell|,p}(r) a_{|\ell|,p}^{(\sigma)}(k\hat{z}) e^{i(kz+\ell\phi)} - H.c. \right], \quad (3)$$

where $\Omega = i\left(\hbar c k \varepsilon_0 / 2 A_{\ell,p}^2 V\right)^{1/2}$ is the normalization constant for LG modes, $e_i^{(\sigma)}\left(b_i^{(\sigma)}\right) = \sqrt{2}^{-1}\left(\hat{x}(\hat{y}) \pm i\sigma \hat{y}(\hat{x})\right)_i$ represents the electric (magnetic) polarization vector for circularly polarized light with $\sigma = \pm 1$ the helicity, with $V$ the quantization volume; $a_{|\ell|,p}^{(\sigma)}(k\hat{z})$ is the annihilation operator; $\exp i(kz + \ell\phi)$ is the phase; and H.c. stands for Hermitian conjugate. The terms that depend on $\hat{x}$ and $\hat{y}$ are the transverse components of the fields while those that depend on $\hat{z}$ are the longitudinal components. 2D structured light possesses only the transverse components, whilst 3D structured light also has the longitudinal component [7]. $f_{|\ell|,p}(r)$ is a radial distribution function given as

$$f_{|\ell|,p}(r) = \frac{C_p^{|\ell|}}{w_0}\left(\frac{\sqrt{2}r}{w_0}\right)^{|\ell|} e^{-\frac{r^2}{w_0^2}} L_p^{|\ell|}\left[\frac{2r^2}{w_0^2}\right], \quad (4)$$

Where $C_p^{|\ell|} = \sqrt{2p!/\pi(p+|\ell|)!}$ is a normalization constant, $w_0$ is the minimum transverse extent of the beam around $z = 0$ (beam waist), and $L_p^{|\ell|}$ is the generalized Laguerre polynomial, $p$ being the radial index. The potential energy responsibe for the optical gradient trapping force originates in forward Rayleigh scattering. In photonic terms, an input laser photon is annihilated at the particle and then an identical photon (same mode) is created at the same particle. This two photon interaction requires second-order perturbation theory to yield the leading order contribution to the potential energy [1]:

$$U = \text{Re} \sum_R \frac{\langle I|H_{\text{int}}|R\rangle\langle R|H_{\text{int}}|I\rangle}{E_R - E_I}, \quad (5)$$

where the initial state of the system is given by $|I\rangle = |E_0(\xi); n(k,\sigma,\ell,p)\rangle$: the particle is in the ground state and the radiation field consists of $n$ photons in the mode $(k,\sigma,\ell,p)$; the final state of the system is identical to the initial; the virtual intermediate state is given by $|R\rangle = |E_\alpha(\xi); (n-1)(k,\sigma,\ell,p)\rangle$. Working strictly in the dipole regime the potential energy of a particle is then a sum of three distinct contributions:

$$U \simeq U_{\text{E1E1}} + U_{\text{E1M1}} + U_{\text{M1M1}}. \quad (6)$$

Chiral effects dependent on material handedness originate in the E1M1 term and so from now on we neglect the pure E1E1 and M1M1 effects as they are independent of material chirality [37]. **Circular polarized input** Using Eq. (1) with the fields Eqs. (2) and (3) in Eq. (5) we produce the following potential energy for a circularly polarized input

$$U_{\text{E1M1}}^{\text{Circ}} = -\frac{I}{2\varepsilon_0 c^2} \text{Re}\left[\left\{\overline{e}_i^\sigma f - \frac{i}{k}\left(f' - \frac{\ell\sigma}{r}f\right)e^{-i\sigma\phi}\hat{z}_i\right\}\right.$$
$$\left. \times \left\{b_j^\sigma f + \frac{1}{k}\left(\sigma f' - \frac{\ell}{r}f\right)e^{i\sigma\phi}\hat{z}_j\right\} G_{ij} + c.c.(i \leftrightarrow j)\right], \quad (7)$$

where we have dropped most dependencies for notational clarity, $c.c.(i \leftrightarrow j)$ stands for taking the complex conjugate and swapping the indices of the expression in brackets, $I = n\hbar c^2 k / A_{\ell,p}^2 V$ is the beam intensity, $f' = \partial_r f_{|\ell|,p}(r)$, and the imaginary quantity referred to as the mixed electric-magnetic polarizability tensor (which has different signs for each chiral particle of an enantiomeric pair) is given explicitly as

$$G_{ij}(\omega,-\omega) = \sum_\alpha \left(\frac{\mu_i^{0\alpha} m_j^{\alpha 0}}{E_{\alpha 0} - \hbar\omega} + \frac{m_j^{0\alpha} \mu_i^{\alpha 0}}{E_{\alpha 0} + \hbar\omega}\right), \quad (8)$$

where the transition dipole moments are defined as $\mu_i^{0\alpha}(m_i^{0\alpha}) = \langle E_0|\mu_i(m_i)|E_\alpha\rangle$ and $E_{\alpha 0} = E_\alpha - E_0$. To account for the randomly oriented nature of chiral molecules in the liquid or gas phase the potential enery (7) must be rotationally averaged. Using well known methods [34] in which each molecule is decoupled from the space-fixed frame into the laboratory frame we produce

$$\langle U_{\text{E1M1}}^{\text{Circ}}\rangle = -\frac{I}{3\varepsilon_0 c^2}\left[\sigma f^2\right.$$
$$\left. + \frac{1}{2k^2}\left(\sigma f'^2 - \frac{2\ell}{r}ff' + \frac{\sigma\ell^2}{r^2}f^2\right)\right]\text{Im} G_{\lambda\lambda}, \quad (9)$$

The optical force $\mathbf{F}$ can be calculated from the energy shift via the well-known relation: $\mathbf{F} = -\text{Re}\nabla U$, its explicit form is given in Supplement 1. The potential energy (9) is plotted in Fig. 1 with the forces $\mathbf{F}$ overlaid for the parallel $\ell = |\sigma|$ and antiparallel $\ell = -|\sigma|$ cases (further plots can be found in Supplement 1).

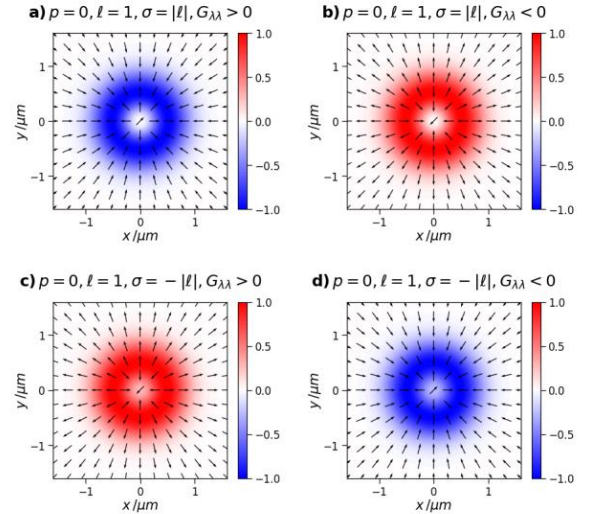

**Fig. 1** Individually normalized enantioselective potential energy (9) with forces overlaid (arrows indicate only direction) $w_0 = \lambda$ in all plots.

This enantioselective trapping mechanism is quite similar to that of circularly polarized plane waves or 2D structured light, except in our system we have the spin-orbit-interactions (SOI) which occur in 3D fields [38] and these influence the helicity density distributions. For example, in the antiparalell case we produce an on-axis helicity density for $|\ell| = 1$.

**Polarization independent enantioselective force** While the physics of the mechanism in the previous section is similar to that of previous chiral gradient forces, we will now highlight a highly distinct enantioselectice *polarization independent* trapping mechanism. Optical helicity density $h$ for propagating plane waves, 2D structured fields, and evanescent waves is proportional to the degree of circular polarization of the input beam $h \propto \sigma$, being zero for linear/randomly polarized inputs and taking on its maximum value for circular polarization. For linearly/unpolarized plane waves there is no possibility for enantioselective gradient forces because $\sigma = 0$ and thus their optical helicity $h = 0$. In comparison, 3D structured LG modes are known to possess a non-zero optical helicity density for linear polarizations [36,39] and it has recently been shown that remarkably it is independent of polarization, being non-zero for even unpolarized light [40]. The rotationally averaged gradient trapping potential energy with an unpolarized 3D LG input may be calculated by averaging (9) over the two orthogonal polarizations $\sigma = 1$ and $\sigma = -1$ (see Supplement 1 for further details):

$$\left\langle U_{\text{E1M1}}^{\text{ind}} \right\rangle = \frac{I}{3\varepsilon_0 c^2} \frac{\ell}{k^2 r} f\!f \, \text{Im} \, G_{\lambda\lambda}. \tag{10}$$

The gradient forces acting on a chiral particle stemming from this polarization independent potential energy are calculated as

$$\left\langle F_{\text{E1M1}}^{\text{ind}} \right\rangle = \hat{r} \frac{I}{3\varepsilon_0 c^2} \left(\frac{C_p^{|\ell|}}{w_0}\right)^2 \left(\frac{\sqrt{2}r}{w_0}\right)^{2|\ell|} e^{-\frac{2r^2}{w_0^2}} \frac{8\ell}{k^2}$$

$$\times \left[ \left(\frac{|\ell|}{w_0^2 r} - \frac{r}{w_0^4}\right) L_p^{2|\ell|} + \left(\frac{2|\ell|}{w_0^2 r} - \frac{4r}{w_0^4}\right) L_p^{|\ell|} L_{p-1}^{|\ell|+1} \right.$$

$$\left. - \frac{2r}{w_0^4} \left( L_{p-1}^{2|\ell|+1} + L_p^{|\ell|} L_{p-2}^{|\ell|+2} \right) \right] \text{Im} \, G_{\lambda\lambda}. \tag{11}$$

The potential energy (10) with the forces (11) overlaid at the focal plane $z = 0$ are shown in Fig. 2. What the graphs show is that for a given sign of $\ell$, the right-handed and left-handed chiral particles in an enantiomeric mixture are subject to discriminatory radial trapping forces which act to separate them into distinct rings in the transverse plane. For example, in Fig. 2a the enantiomer with $G > 0$ is pushed away from the beam axis and towards the blue (outer) ring, meanwhile in 2b it shows the $G < 0$ is pushed towards the central spot and away from the outer ring. The results are dependent on the sign of $\ell$, and keeping everything else the same, for $-\ell$ the graphs in Fig 2. are the same but with the signs reversed.

For example, for $-\ell$ the 2a $G > 0$ would look exactly like 2b.
**Discussion** The chiral forces derived here using perturbative QED can alternatively be interpreted as the force arising from the gradient of the optical helicity density $h$ [27].

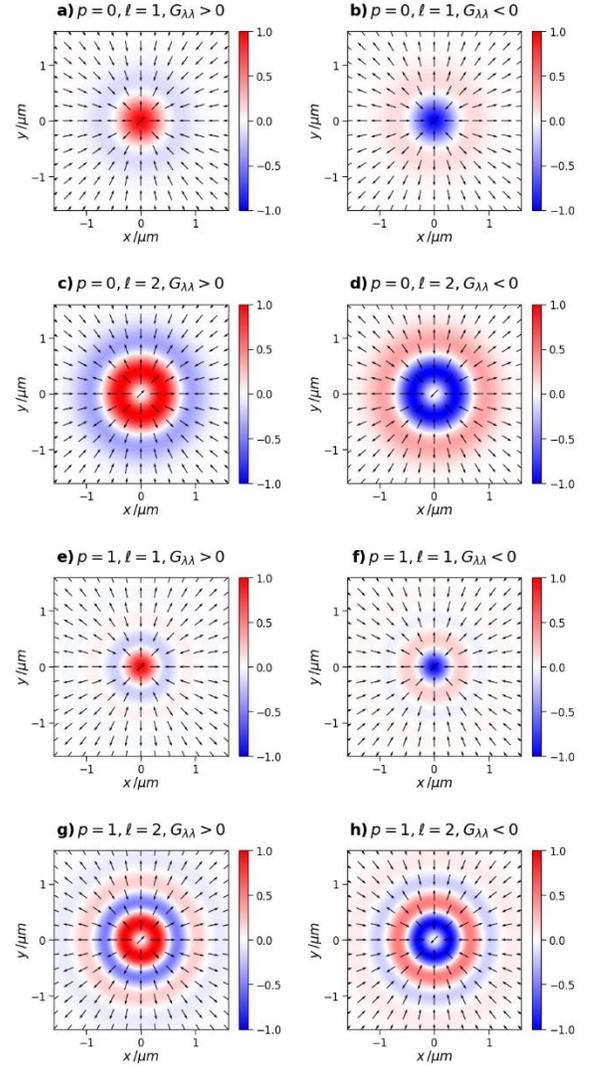

**Fig. 2** Individually normalized polarization independent enantioselective potential energy (10) with vector optical force (11). Otherwise as Fig. 1.

It is clear that the forces in Figs. 1 and 2 are proportional to $(\text{Im} \, G_{\lambda\lambda}) \nabla h$, where $h$ is given by Eqs. (14) and (17) in Ref. [36], respectively. We note that our *conservative* discriminatory trapping force derived using QED relates to the $-\omega 4\pi \, \text{Re}\{\chi\} \nabla h$ term derived with a classical field in Eq. 2 of Ref [27]. For the propagating plane wave and evanescent wave modelled in that study and elsewhere [41] the helicity density is $h \propto \sigma$ and so for linear/randomly polarized propagating plane or evanescent waves conservative enantioselective forces analogous to our 3D structured vortex polarization independent forces do not exist. A key difference between evanescent fields and propagating plane waves is that the former possess longitudinal field components. The lateral *non-conservative* optical forces presented in Ref. [27] are due to transverse spin angular momentum density [42] in evanescent waves. Although the evanescent fields have polarization independent transverse spin density [43], they possess canonical momentum density only in the direction of propagation, and so projecting the transverse spin onto the momentum density in evanescent waves produces zero helicity. We have highlighted a similar chirality-sorting force as that discovered by Hayat et al [27] but ours is a conservative gradient force in 3D optical vortex tweezing systems, its magnitude is proportional to the tightness of the focus, but also increases for higher $p$ values as it stems from the gradients of the field. Evidently for natural chiral particles (e.g. molecules) the magnitudes of the forces involved will be small due their small $G_{\lambda\lambda}$, however they are considered experimentally distinguishable [1], though are yet to be observed. Experimental realization of the mechanisms outlined here therefore have more potential in systems comprising of nanoparticles or the use of plasmonic enhancement in nanophotonic setups [19,44,45]. This work has provided further evidence of the significant potential of 3D structured light in light-matter interactions and nanophotonics [7]. To be clear, while enantiomer separation schemes utilizing the optical helicity of circularly polarized plane waves in gradient force mechanisms have been put forward [13,15], linear/randomly polarized propagating plane waves possess no longitudinal fields or optical helicity and so the polarization independent discriminatory force (11) could never have been envisaged under the plane wave approximation, nor a 2D structured light field.

**Funding.** Leverhulme Trust (Grant Number ECF-2019-398).

**Data Availability.** Data underlying the results presented in this Letter are not publicly available at this time but may be obtained from the authors upon reasonable request.

**Disclosures**. The author declares no conflicts of interest.

**Supplemental document.** See Supplement 1 for supporting content.